\documentclass[prb,
                    a4,twocolumn
                  ] {revtex4-1}
\usepackage{epsfig}
\usepackage{subfigure}
\def\e{\begin{equation}}
\def\f{\end{equation}}

\begin{document}

\title{Bistatic scattering characterization of a three-dimensional broadband cloaking structure}

\author{Pekka Alitalo$^{1,2}$}
%\email{pekka.alitalo@aalto.fi}

\author{Ali E. Culhaoglu$^2$}

\author{Andrey V. Osipov$^2$}

\author{Stefan Thurner$^2$}

\author{Erich Kemptner$^2$}

\author{Sergei A. Tretyakov$^1$}

\affiliation{$^1$~Department of Radio Science and Engineering / SMARAD Centre of Excellence,\\ Aalto University, P.O. Box 13000, FI-00076 Aalto, Finland\\
{\rm E-mail: pekka.alitalo@aalto.fi}\\ $^2$~Microwaves and Radar
Institute, German Aerospace Center DLR, 82234 Wessling,
Germany\vspace{1.0cm}}

\date{October 18, 2011}

\begin{abstract}

Here we present the results of full experimental characterization
of broadband cloaking of a finite-sized metallic cylinder at
X-band. The cloaking effect is characterized by measuring the
bistatic scattering patterns of uncloaked and cloaked objects in
free space and then comparing these with each other. The results
of the measurements demonstrate a broadband cloaking effect and
are in good agreement with numerical predictions.\newline\newline

%PACS numbers: 41.20.-q, 41.20.Jb, 42.25.Bs

\end{abstract}

\maketitle

\section{Introduction}

The concept of electromagnetic cloaking has been found by many as
an exciting example of the power of artificial electromagnetic
materials (metamaterials). The concepts of transformation
optics~\cite{Leonhardt_Science,Pendry_Science} and scattering
cancellation~\cite{Alu_transparency,Alu_review} describe different
ways of making scattering objects invisible, that is, reducing the
total scattering cross section of an object ideally to zero. First
experimental proofs of reduction of the total scattering from
specific objects with these techniques have been provided by
measuring the microwave field distributions near cloaked and
uncloaked objects.\cite{Alu_measurement,Smith_measurement}

Recently, some alternative ways of achieving electromagnetic
cloaking by employing waveguiding structures instead of complex
materials have been
proposed.\cite{Alitalo_review1,Alitalo_review2} These structures
are composed of simple transmission lines or specifically shaped
metallic structures that enable the electromagnetic wave either to
go through the object~\cite{Alitalo_TAP} or to go around the
object.\cite{MPcloak_PRL,MPcloak_PRB} Both these cloaking methods
have been already studied numerically and experimentally. However,
the experimental results for these structures were obtained in a
rectangular waveguide setup, which does not allow the analysis of
the total scattering cross sections or total scattering widths.

In this paper we describe microwave measurements at X-band of the
bistatic scattering patterns of a finite-sized waveguiding cloak
structure composed of stacked conical metal plates (therefore also
called the ``metal-plate cloak'').\cite{MPcloak_PRL,MPcloak_PRB}
The cloaking effect is experimentally characterized by measuring
the bistatic scattering patterns and comparing the total
scattering widths of both uncloaked and cloaked objects. In
addition to bistatic measurements, we employ the forward
scattering theorem~\cite{FWtheorem} to evaluate the total
scattering width using only the fields scattered in the forward
direction. It is shown that the both approaches agree well with
the numerical results.

It should be noted that the previous experimental demonstrations
of reduction of total scattering widths by cloaking
devices~\cite{Alu_measurement,Smith_measurement} have been
obtained by emulating infinitely long structures, since in both
cases the measurements were carried out in a parallel-plate
waveguide instead of free space. Here we carry out the
measurements in free space conditions, using two antennas to
measure the scattered fields produced by finite-sized objects. In
addition to measurements we also present the corresponding
numerical results for the total scattering widths and bistatic
scattering patterns that are obtained with the commercial
numerical software ANSYS HFSS.\cite{HFSS}

\section{Geometry and dimensions of the cloak structure}

\begin{figure}[t!]
\centering \epsfig{file=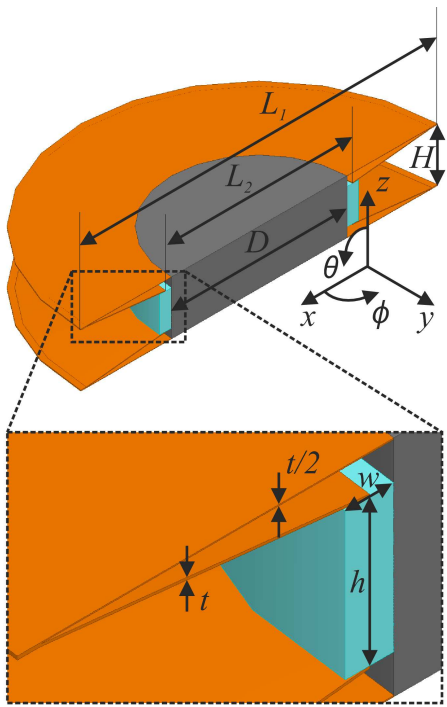, width=0.33\textwidth}
\caption{(Color online) Geometry of the cloak enclosing the
cloaked object. The cloaked object is a solid cylinder with
diameter $D$. One half of a single unit cell of the cloak, cut
along the $xz$-plane, is shown.}\label{cloak}
\end{figure}

The cloak geometry is the same as presented in Ref.~11, see
Fig.~\ref{cloak}. The conical metal plates, realized from and
modelled as copper, form a periodic structure that surrounds a
metal cylinder which is the object to be cloaked. The plates form
a set of waveguides into which the electromagnetic wave with the
electric field parallel to the axis of the cylinder can couple to.
The wave travels inside the structure, with very weak reflections,
around the cylinder placed in the center. Essentially this
cloaking device is designed to work for objects that are
electrically small since the wave inside the cloak must travel a
distance larger than the wave that travels in free space along a
straight line. However, it has been shown\cite{MPcloak_PRB} that
the diameter of the cloaked cylinder can be as large as about one
wavelength without much compromising the cloaking effect.
Therefore we use the structure and dimensions presented in Ref.~11
to demonstrate the cloaking of a metal (brass) cylinder having the
diameter of 30~mm at X-band (specifically, in the band 8.2~GHz --
12.4~GHz). The optimized dimensions for cloaking at frequencies
around 10~GHz are shown in Table~I. A dielectric support of height
$h$ and thickness $w$ has been added between two adjacent metal
plates to enable practical realization of the structure. This
dielectric is Rohacell$^{\textregistered}$ 51 HF\cite{Rohacell}
and is numerically modelled as having the relative permittivity of
$\epsilon_r=1.07$ and loss tangent of $\tan{\delta}=0.003$. These
material properties are so close to the free space values that the
layers do not practically affect the cloak's operation, as can be
concluded by comparing the numerical results shown later in this
paper with the results presented in Ref.~11.

\begin{table}[t!]
\centering \caption{Dimensions of the cloak operating at around
10~GHz~\cite{MPcloak_PRB}.} \label{table1}
\begin{tabular}{|c|c|c|c|c|c|c|} \hline
$D$ & $H$ & $h$ & $L_1$ & $L_2$ & $t$ & $w$\\

\hline

30~mm & 9.2~mm & 6~mm & 61~mm & 32~mm & 0.1~mm & 2~mm\\

\hline
\end{tabular}
\end{table}

\section{Experimental setup for measuring the bistatic scattering}

\begin{figure}[t!]
\centering \epsfig{file=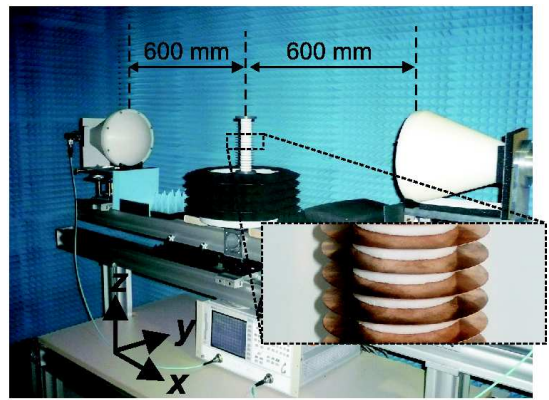, width=0.4\textwidth}
\caption{(Color online) Photograph of the measurement setup at
DLR, including the scatterer to be measured (in the center). The
inset shows a magnification of a part of the cloak.}\label{photo}
\end{figure}

We use a bistatic measurement setup composed of two antennas to
measure the fields scattered by uncloaked and cloaked objects in
various directions in the $xy$-plane. The transmitting antenna is
fixed at $\phi=0^{\circ}$ while the receiving antenna is swept
from $\phi=22^{\circ}$  to $\phi=180^{\circ}$. Both antennas are
operating with the vertical polarization (electric field parallel
to the $z$-axis). See Fig.~\ref{photo} for a photograph of the
setup. The transmitted and received signals are measured
with a vector network analyzer (Agilent HP 8719D) in the band
8.2~GHz-12.4~GHz with 401 frequency points. Both antennas are
equipped with microwave lenses that focus the antenna beams exactly
on the object that is measured. The beam's
half-power width at the focus is approximately 45~mm and the
measured structures are made considerably higher than the
beamwidth to avoid any scattering effects caused by the finite
height. The cloak to be measured is made of 20 unit cells shown in
Fig.~\ref{cloak}, i.e., the cloak (and the cloaked cylinder) has
the height 184~mm.

Under these conditions the bare cylinder and the cloak scatter
mostly in the $xy$-plane and therefore we can characterize the
cloaking effect using the total scattering width instead of the
total scattering cross section. We measure the scattering pattern
of the cloaked/uncloaked object for angles
$22^{\circ}-180^{\circ}$ since the scattering structures are
symmetrical with respect to the $xz$-plane and the measurement of
angles $0^{\circ}-22^{\circ}$ is not possible due to the finite
size of the antennas. However, the lack of data for angles
$0^{\circ}-22^{\circ}$ will not be problematic since most of the
scattering occurs in the forward direction (close to
$\phi=180^{\circ}$), and we can compare our measured results with
numerical results in the same angular range. All the measurements
presented in this paper have been carried out with an angular step
of $0.5^{\circ}$.

%\begin{figure}[t!]
%\centering \epsfig{file=setup.eps, width=0.33\textwidth}
%\caption{Illustration of the measurement setup. The RX antenna can be rotated around the origin from 22$^{\circ}$ to 180$^{\circ}$.}\label{setup}
%\end{figure}

For every angle we measure the complex scattering parameter of
the studied object ($S_{\rm 21,O}$). In addition, the
measurement is repeated without the scatterer ($S_{\rm 21,FS}$,
i.e., transmission in free space) to account for the crosstalk
between the transmitting and receiving antennas. A value directly
proportional to the scattered electric field is then obtained from
these two measurements with

\begin{equation}
E_{\rm sca,O}(f,\phi) = S_{\rm 21,O}(f,\phi) - S_{\rm
21,FS}(f,\phi),
\end{equation} where $f$ is the frequency and $\phi$ is the angle in the $xy$-plane.

One of the most practical ways to characterize a cloak and to
determine the efficiency of cloaking is to study the total
scattering width of the cloaked object, normalized by the total
scattering width of the uncloaked object. The same analysis has
been used in Ref.~5. The scattered field intensities are
integrated over the $xy$-plane for both objects. The normalized
total scattering width ($\sigma_{\rm W,norm}$, i.e., the total
scattering width of object 1 normalized by the total scattering
width of object 2) is therefore

\begin{equation}
\sigma_{\rm W,norm}(f) = \frac{\int{|E_{\rm
sca,O1}(f,\phi)|^2}d\phi}{\int{|E_{\rm sca,O2}(f,\phi)|^2}d\phi},
\end{equation} where ``O1'' stands for object 1 and ``O2'' for object 2.

All our scattering objects are cylindrically symmetric, so it
would be enough in (2) to integrate only from $0^{\circ}$ to
$180^{\circ}$, but since we are limited in the angular range of
the measurement setup, we have to approximate (2) by integrating
from $22^{\circ}$ to $180^{\circ}$. However, the backscattered field at $\phi=0^{\circ}$ can be determined by measuring the reflection coefficient ($S_{\rm 11}$) of the transmitting antenna and equating it to the scattered field. We are
not using this value of the scattered field in the integration,
but we can use it for making sure that the scattering objects
behave as expected also in the back direction.

%\begin{equation}
%E_{sca}(\phi=180) = S_{11}.
%\end{equation}

The forward scattering (or optical) theorem~\cite{FWtheorem}
states that the total scattering width (or cross section) is
directly related to the complex field scattered in the forward
direction. This theorem offers a convenient way to estimate the
total scattering widths of the objects studied here since only the
scattered fields at $\phi=180^{\circ}$ need to be considered. To
employ the theorem, we need to find the far-field scattering
coefficients~\cite{FWtheorem} of the studied objects. These are found by using a calibration scatterer,
such as a metal cylinder. Specifically, the total scattering width
of an object is found with

\begin{equation}
\sigma_{\rm W,FW}(f) = -\frac{4}{k}{\rm Re}\left\{
P_{cal}\frac{E_{\rm sca,O}(f,\phi=180)}{E_{\rm
sca,cal}(f,\phi=180)} \right\},
\end{equation} where $k$ is the wavenumber, $P_{cal}$ is the analytically known~\cite{FWtheorem} far-field scattering coefficient of the calibration object, and $E_{\rm sca,O}$, $E_{\rm sca,cal}$ are the measured
scattered electric fields of the studied object and calibration
object, respectively. The normalized total scattering width is
obtained by taking the ratio of $\sigma_{\rm W,FW}$ of two
different objects.

%We employ the forward scattering theorem to estimate the normalized total scattering widths by first calibrating the
%scattered fields $E_{sca}(\phi=0)$ with scattering from a known object. This we do by measuring a metal cylinder and then obtaining a
%calibration vector by equating the measured scattered fields from this object to it's analytical value. The normalized total scattering width is then obtained from

\begin{figure}[t!]
\centering \epsfig{file=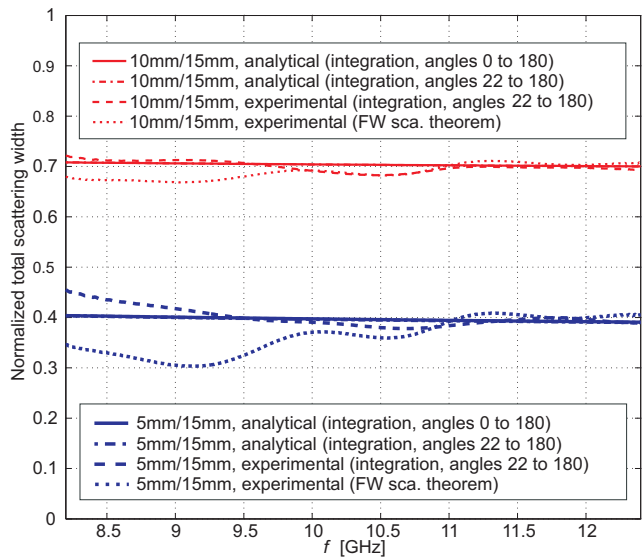, width=0.47\textwidth}
\caption{(Color online) Analytical and experimental results for
normalized total scattering widths. Total scattering widths of
cylinders with radii 5~mm and 10~mm are normalized to the total
scattering width of a cylinder having the radius
15~mm.}\label{cylcomparison}
\end{figure}

We first verify the operation of the measurement setup by
measuring metal cylinders of different diameters so that we can
compare the measured scattering widths with the known analytical
results. In this way we can estimate the error which is introduced
by integrating in (2) only over the limited range of angles. We
measure three different metal cylinders with radii 5~mm, 10~mm,
and 15~mm (the height of all cylinders equals 184~mm). Then we use
(1) for the scattered fields and (2) to obtain the normalized
total scattering widths using the integration method. We also
employ the forward scattering theorem (3), where we use the
cylinder with radius 10~mm as the calibration object.
Fig.~\ref{cylcomparison} presents analytical and experimental
results for the normalized total scattering widths of various
cylinders. The measured results agree well with the analytical
ones, although a little bit of fluctuation is naturally seen in
the measured values. It should be noted that two analytical
results are shown for comparison: the exact solution of (2), and
one obtained by integrating in (2) from $22^{\circ}$ to
$180^{\circ}$. These analytical curves overlap and with the
scaling in the plot it is impossible to distinguish these results
from each other. This is expected since the approximation done in
the integration disregards only a small portion of the angular
range, and moreover, the objects scatter much more in the forward
direction than in the back direction. One important factor is also
that close to the back direction (in $0^{\circ}$ to $22^{\circ}$),
the scattering cross section is stable with respect to the angle.
Also the forward scattering theorem gives a reasonable result,
although we can conclude that the integration method gives a more
accurate result. The reason for this is that the scattered field
in the forward direction ($\phi=180^{\circ}$) has larger
measurement error than the scattered fields in other directions.
This is very much expected since for small scatterers and for
angles close to $180^{\circ}$, the  parameters $S_{\rm
21,O}(f,\phi)$ and $S_{\rm 21,FS}(f,\phi)$ are at maximum and
close to each other which leads to the larger error.

\section{Experimental and numerical results of cloaking}

\begin{figure}[t!]
\centering \epsfig{file=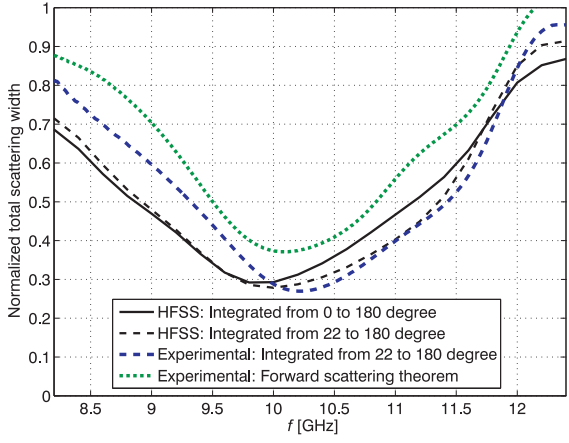, width=0.49\textwidth}
\caption{(Color online) Experimental and numerical results for the
total scattering width of the cloaked object, normalized to the
total scattering width of the uncloaked
object.}\label{TSW}\vspace{1.0cm}
\end{figure}

A cloak with 20 unit cells of Fig.~\ref{cloak} was assembled on a
metal cylinder having the diameter 30~mm. The cloaked and
uncloaked cylinders were then measured as described above.
Fig.~\ref{TSW} presents the normalized total scattering width for
the cloak studied in this paper, demonstrating that the cloak
reduces the total scattering width of the uncloaked metal cylinder
by about 70\% in maximum. The wide operation bandwidth is also
demonstrated, since the cloak is shown to reduce the total
scattering width of the cylinder by more than 50\% in a relative
frequency bandwidth of about 20\%. The center of the cloaking band
is at around 10~GHz, as predicted by numerical results. It should
be emphasized that based on the results of
Fig.~\ref{cylcomparison}, the integration method is expected to be
more accurate than that based on the forward scattering theorem.
However, it is clear that the both curves demonstrate the cloaking
effect.

\begin{figure}[t!]
\centering \subfigure[]{\epsfig{file=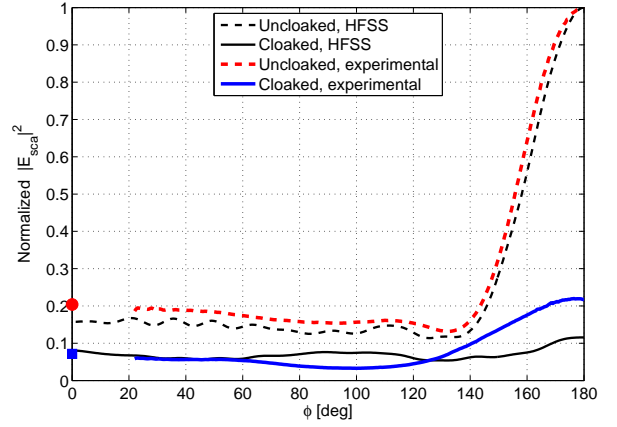,
width=0.49\textwidth}} \subfigure[]{\epsfig{file=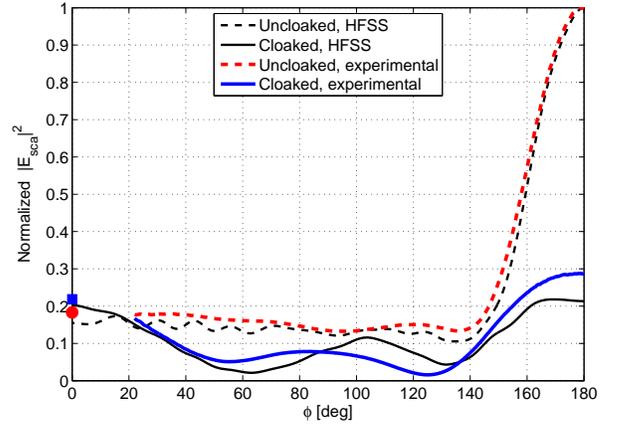,
width=0.49\textwidth}} \subfigure[]{\epsfig{file=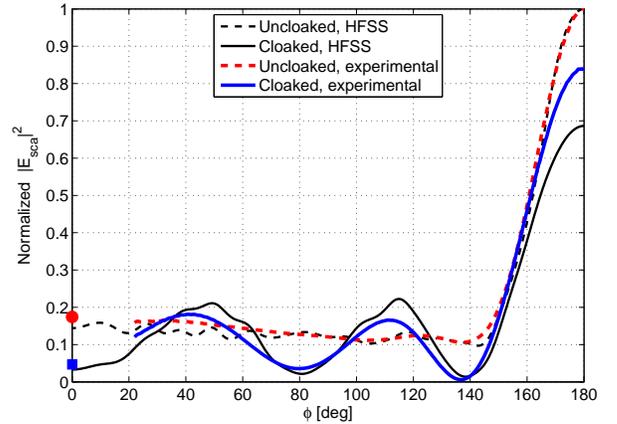,
width=0.49\textwidth}} \caption{(Color online) Angular dependency
of the scattered field intensities, normalized to that of the
uncloaked object at $\phi=180^{\circ}$. (a)~$f=10$~GHz,
(b)~$f=11$~GHz, (c)~$f=12$~GHz. The results of the monostatic
measurements ($\phi=0^{\circ}$) are shown as squares (cloaked) and
circles (uncloaked).}\label{angular}
\end{figure}

Although the most important figure of merit for the cloak is the
normalized total scattering width, it is interesting to look at
the angular dependence of the bistatic scattering patterns
at fixed frequencies. Fig.~\ref{angular} presents the normalized scattered field intensities as a function of $\phi$ for frequencies
10~GHz, 11~GHz, and 12~GHz. It is obvious that the experimental
results differ somewhat from the numerical ones, but the numerical
and experimental results for the overall performance of the cloak
are in good agreement.

%Especially at the higher frequencies the angular dependence of the scattered field intensities are similar in the numerical and experimental results.

%For example, the normalized scattered field intensity in the forward direction [i.e., $E_{sca,cloaked}(f,\phi=0)^2/E_{sca,uncloaked}(f,\phi=0)^2$ ]
%is presented in Fig.~\ref{FSW}, showing satisfactory agreement between the numerical and experimental results.

%It is clear that in the experiment, the
%cloak scatters slightly more in the forward direction than in anticipated by the simulations. This can be explained by the inaccuracy of the manual assembly of the cones.
%By studying the scattering behaviour of the cloak with introducing certain non-idealities, we have found that even a slight (less than a millimeter) introduction of asymmetry in
%the cloak (so that the height of the cones in Fig.~\ref{cloak} is not exactly the same) can introduce such an increase in the forward scattered field.

%However, since the normalized total scattering widths of Fig.~\ref{TSW} show actually better performance for the experiment than for the simulation, we can expect that
%the realized cloak works somewhat better than expected in directions other than $\phi=0^{\circ}$.

\section{Conclusions}

Experimental confirmation of scattering reduction by a finite-sized
metal-plate cloak operating in the X-band has been presented. A bistatic
free space measurement setup has been established for this purpose. The
efficiency of the cloak has been demonstrated by determining the total
scattering widths of an uncloaked and cloaked metallic cylinder of finite height.
Measurement results have been compared with numerical ones. The results are in good
agreement with each other. The cloak has been shown to reduce the total
scattering width of the cylinder by more than 50\% in a relative frequency bandwidth of about 20\% around 10~GHz.

%We have presented experimental confirmation of scattering reduction by
%a finite-sized metal-plate cloak operating in the X-band. The
%results are obtained with a bistatic measurement setup with which
%we have been able to measure the fields scattered by uncloaked and
%cloaked objects in various directions. By taking into account the
%scattering in all the measured directions, we have presented an
%approximation of the total scattering width of the cloaked object,
%normalized to the total scattering width of the uncloaked object,
%and demonstrated the cloaking efficiency at various frequencies.
%In addition, we have employed the forward scattering theorem to
%estimate the normalized total scattering width by using only the
%complex scattered field in the forward direction. The experimental
%results have been compared with numerical ones resulting in good
%agreement with each other. In addition, we have studied the
%scattering behaviour in various directions at specific frequencies
%to confirm that the realized cloak is working as expected based on
%the numerical models.

%\section*{Acknowledgments}

This work has been partially funded by the Academy of Finland and Nokia
through the center-of-excellence program. The work of P. Alitalo has been supported by the Academy of Finland through post-doctoral project funding. P. Alitalo acknowledges the work of Mr. Eino Kahra in helping in the manufacturing of the cloak.

\end{document}